\documentclass[fleqn,12pt,twoside]{article}
\usepackage{espcrc1}
\usepackage{graphicx}%

\usepackage{epsfig}
\title{On possible existence of the dibaryon resonance $d^\star_1$
and its role in the $np\gamma$ and $pd\gamma$ processes below the pion threshold.}

\author{ A.S.Khrykin\address
{Joint Institute for Nuclear Research, Dubna, 141980 Russia}}
\begin{document}
\maketitle
\begin{abstract}
We give reasons for the existence of the $NN$ decoupled dibaryon
resonance $d^\star_1$(1956). Strong
evidence for its presence has first been found in
the energy spectrum of coincident photons emitted at $\pm 90^0$
from the $pp \to \gamma \gamma X$
process at 216 MeV measured by the DIB2$\gamma$
collaboration at JINR. As further experimental indications of
the $d^\star_1$(1956) existence we present those found in
the available photon energy spectra of $np\gamma$,
$pd\gamma$, and $pA\gamma$ reactions below the pion threshold.
It is noted that serious discrepancies between
the $np\gamma$ and $pd\gamma$ experimental cross sections
and  theoretical calculations can reasonably be explained by the fact
that the latter did not take into account the $d^\star_1$ effect.
\end{abstract}
\section{Introduction}
Recently the $DIB2\gamma$ collaboration at JINR, carrying out
searches for the $NN$ decoupled dibaryon resonances
in the reaction $pp \to \gamma\gamma X$ at an energy below the
$\pi NN$ threshold \cite{GerKh93,EGK95}, has reported \cite{PRC64} the observation of
such a state with a mass of about 1956 MeV, which was called $d^\star_1$.
The conclusion for the existence of this resonance was drawn from
the analysis of the energy spectrum for energetic ($E_\gamma \geq 10$ MeV)
coincident photons emitted at angles of $\pm 90^0$ in the laboratory
system from the reaction $pp \to \gamma\gamma X$ at an energy
of about 216 MeV. The measured spectrum shown in Fig.1 consists of a
narrow peak at a photon energy of about 24 MeV and
a relatively broad peak in the energy range (45 - 75) MeV.
The statistical significances for the
narrow and broad peaks are 5.3$\sigma$ and 3.5$\sigma$, respectively.
The width of the narrow peak (FWHM) is about 8 MeV,
compatible with that of experimental resolution.
This behavior of the photon spectrum conforms to the signature
of the $NN$ decoupled dibaryon resonance $d^\star_1$(1956)
that is believed to be formed
in the radiative process $pp \to \gamma d^\star_1$ and then to decay via
the $d^\star_1 \to pp \gamma$ mode, thereby contributing to
the $pp \to \gamma \gamma X$ reaction.
The photon energy spectrum calculated with the help of
Monte Carlo simulations under the assumption that the process
$pp \to \gamma \ d^\star_1(1956) \to pp \gamma\gamma$ is
the only mechanism of the $pp \to pp \gamma\gamma$ reaction
proved to be in reasonable agreement with the experimental one.
The differential cross section of the resonance production
of two photons emitted symmetrically at $\theta_{lab} = \pm 90^0$
from the process $pp \to \gamma \ d^\star_1 \to pp \gamma\gamma$ at an energy
of 216 MeV was estimated to be $\sim$ 9 nb/sr$^2$.
The total cross section for the $d^\star_1$(1956) production in
the $pp\gamma\gamma$ reaction at 216 MeV found as a result of multiplying
this differential cross section by $(4\pi)^2$ amounts to about 1.4 $\mu b$.
\begin{figure}[htb]
\begin{minipage}[c]{75mm}
\includegraphics[width=70 mm]{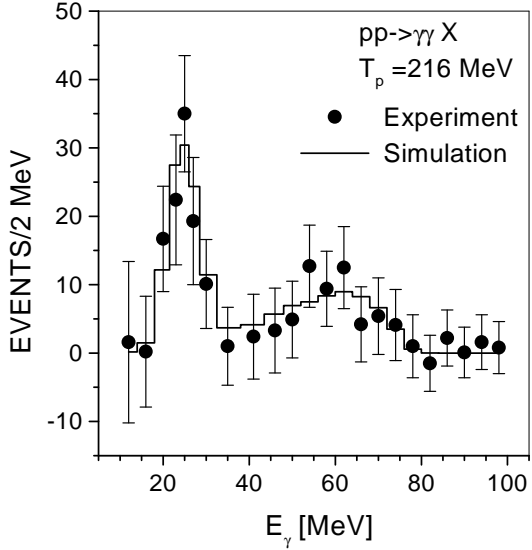}
\end{minipage}
\hspace{\fill}
\begin{minipage}[c]{80mm}
\caption{\small{
Energy spectrum of coincident photons from the $pp \to \gamma\gamma X$
process at 216 MeV measured at 90$^0$ and the spectrum calculated
with the help of Monte Carlo simulation
for the process $pp \to \gamma d^\star_1 \to \gamma\gamma pp$
with a $d^\star_1$ mass of 1956 MeV.}}
\end{minipage}
\vspace*{-0.5 cm}
\end{figure}
The existence of the $d^\star_1$(1956)
resonance implies, in particular, that in addition to the conventional
mechanisms of photon production in $NN$ interactions below the
$\pi NN$ threshold there should exist
a new one, the dibaryon mechanism
$NN \to \gamma d^\star_1 \to NN \gamma \gamma$.
Obviously, depending on kinematical conditions of the $NN\gamma$
and $NA\gamma$ reactions this mechanism would make some contribution
to inclusive photon energy spectra of these reactions.
Such a contribution should be mainly concentrated in two
photon energy regions specific for this dibaryon resonance,
which are determined by energies of photons arising from the $d^\star_1$(1956) formation and decay.
In the center-of-mass system, the energy of a photon connected with
the dibaryon formation ($E_F$)
is determined by its mass $M_R$ and the energy of colliding nucleons
($W=\sqrt{s}$) as $E_F=(W^2-M_R^2)/2W$,
while the energy of a photon associated with the radiative dibaryon
decay ($E_D$) in the rest frame of this resonance is defined as
$E_D=(M_R^2-M_{NN}^2)/2M_R$, where $M_{NN}$ is
the invariant mass of the final $NN$ state populated as a result of this
decay.
The contribution from the dibaryon mechanism to the photon energy
spectrum of the $NN\gamma$ or $NA\gamma$ reaction
would compete with a substantial contribution from the conventional
mechanisms.
 Moreover, in the case of $NA$ collisions even at an energy below
the $\pi NN$ threshold one should have in view the additional contribution
from $\pi^0$ decays produced in $NN$ collisions due to the nucleon Fermi
motion in nuclei.
At the same time, since the bremsstrahlung cross section in $np$ scattering
at energies around 200 MeV is substantially larger than
that for $pp$ scattering,
one can expect that the contribution of the dibaryon mechanism in
$np$ and, hence, in $NA$ scattering is also larger than in $pp$ scattering.

In search of further evidence for the existence of the $d^\star_1$(1956)
dibaryon resonance we decided to analyze the available
experimental data on photon production in $NN$ and $NA$ collisions
at energies below the $\pi NN$ threshold resulting in the photon energy
spectra of the processes explored.
\section{Photon energy spectra of $np\gamma$, $pd\gamma$, and $pA\gamma$ reactions}
The inclusive photon energy spectrum of the reaction $np\to \gamma
X$ was measured at a neutron energy of 170$\pm$35 MeV for high
energy photons ($E_\gamma \ge 20$ MeV) emitted at the angle
$\theta_{lab}=90^0$  \cite{Malek}. This spectrum (data
were taken from Ref. \cite{Malek} and binned in 6 MeV energy
interval to reduce their statistical fluctuations) is shown in
Fig. 2a together with the prediction of the OBE model
\cite{Shafer} folded with a Gaussian with $FWHM=70$ MeV
to allow for an energy spread of an incident beam and
detector resolution. A resonance-like structure in the energy range
from 45 to 75 MeV is clearly seen in this spectrum. At the same
time the calculated photon energy spectrum does not contain even a
hint of the presence of any structure there. To solve the problem of
this structure the authors \cite{Brady} assumed that it was
caused by the contribution from a capture process, that
is, the process $n + p \to d +\gamma$, which was not included in the calculations
\cite{Shafer,Jetter}. But since such a process
(see Fig. 2a) should contribute in an energy range extending
beyond the structure in question, it was suggested in Ref.
\cite{Brady} that the authors \cite{Malek} had incorrectly calibrated
the energy response of their $\gamma$ detector
and so the capture peak was shifted by about 20 MeV.
However, if this is not the case and
at least the shape of the $np\gamma$ spectrum \cite{Malek}
is correct, then the question arises: what is the reason for
this structure origin? In Fig. 2a we also show the energy
spectrum of photons ($E_\gamma \ge$ 20 MeV) emitted at
$90^0_{lab}$ from the process $np \to \gamma d^\star_1(1956) \to \gamma
\gamma np$ at 170$\pm$35 MeV calculated by the Monte Carlo method for
the $d^\star_1$(1956) production cross section of 4.3 $\mu b$. As can
be seen, the $d^\star_1$(1956) contribution to the $np\gamma$
photon spectrum lies in the same energy range as the structure
of interest and therefore the dibaryon mechanism can be considered as
a likely cause for appearance of this structure.
In this connection the existing photon energy spectra of
the $pd \to \gamma X$ \cite{Pinst,Clayt} reaction
are of particular interest.
This is because for low and intermediate photon
energies the $pd\to \gamma X$ spectrum is a good approximation
to the free $np\gamma$ spectrum and thus it should reflect features
inherent in the $np\to \gamma X$ reaction.
The inclusive $pd\gamma$ photon energy spectra were
measured at an incident energy of 200 MeV by the Grenoble
group \cite{Pinst} and at 145 and 195 MeV by the Michigan State
group \cite{Clayt}. The spectra measured by those groups for
$\theta_{lab}=90^0$ at an energy of 200 MeV and 195 MeV are in excellent
agreement. The experimental \cite{Clayt} and calculated
\cite{Nak92} inclusive photon energy spectra of the $pd\gamma$
reaction at 195 MeV and $\theta_{lab}=90^0$ are shown in Fig. 2b.
We see again that the experimental spectrum clearly exhibits a characteristic
structure in the range (45 - 75) MeV, the presence of which leads to
serious disagreement between the experimental and theoretical data.
This fact makes it reasonable to assume that the
structure observed in the energy range
(45 - 75) MeV of both the $np\gamma$ and $pd\gamma$
inclusive photon spectra can arise from the same underlying mechanism. We
believe that such a mechanism is the dibaryon mechanism in
question.
In Fig. 2b we show the dibaryon mechanism contribution to the $pd\gamma$
photon energy spectrum for the $d^\star_1$(1956) production
cross section of 9.4 $\mu b$ calculated by the Monte Carlo method
with the Fermi motion of nucleons in the deuteron taken into account.
Simply adding this contribution to the theoretical spectrum \cite{Nak92}
we find that the agreement between the experimental and theoretical
spectra becomes significantly better.
Finally, the inclusive photon energy spectra were
also measured in $pA$ collisions \cite{ClaytpN,Pinst}.
Obviously, the clearest manifestation of
the dibaryon mechanism should be anticipated in spectra
for light nuclei. In Fig. 2c we show the inclusive photon energy
spectrum for carbon (lighter nucleus) measured at an energy of 200 MeV
and corrected for the $\pi^0$ contribution \cite{Pinst}.
The structure in the energy range (45 - 75) MeV is also seen in this
spectrum, it is very likely to be of the same origin as that observed
in the $np\gamma$ and $pd\gamma$ spectra.
In Fig. 2c we also show the dibaryon mechanism contribution to the
$p+C \to \gamma X$ photon energy spectrum for the $d^\star_1$(1956)
production cross section of 56.5 $\mu b$. We can see that this
contribution is situated in the same energy region as the structure in question.
\begin{figure}[htb]
\begin{minipage}[c]{75mm}
\includegraphics[width=70 mm]{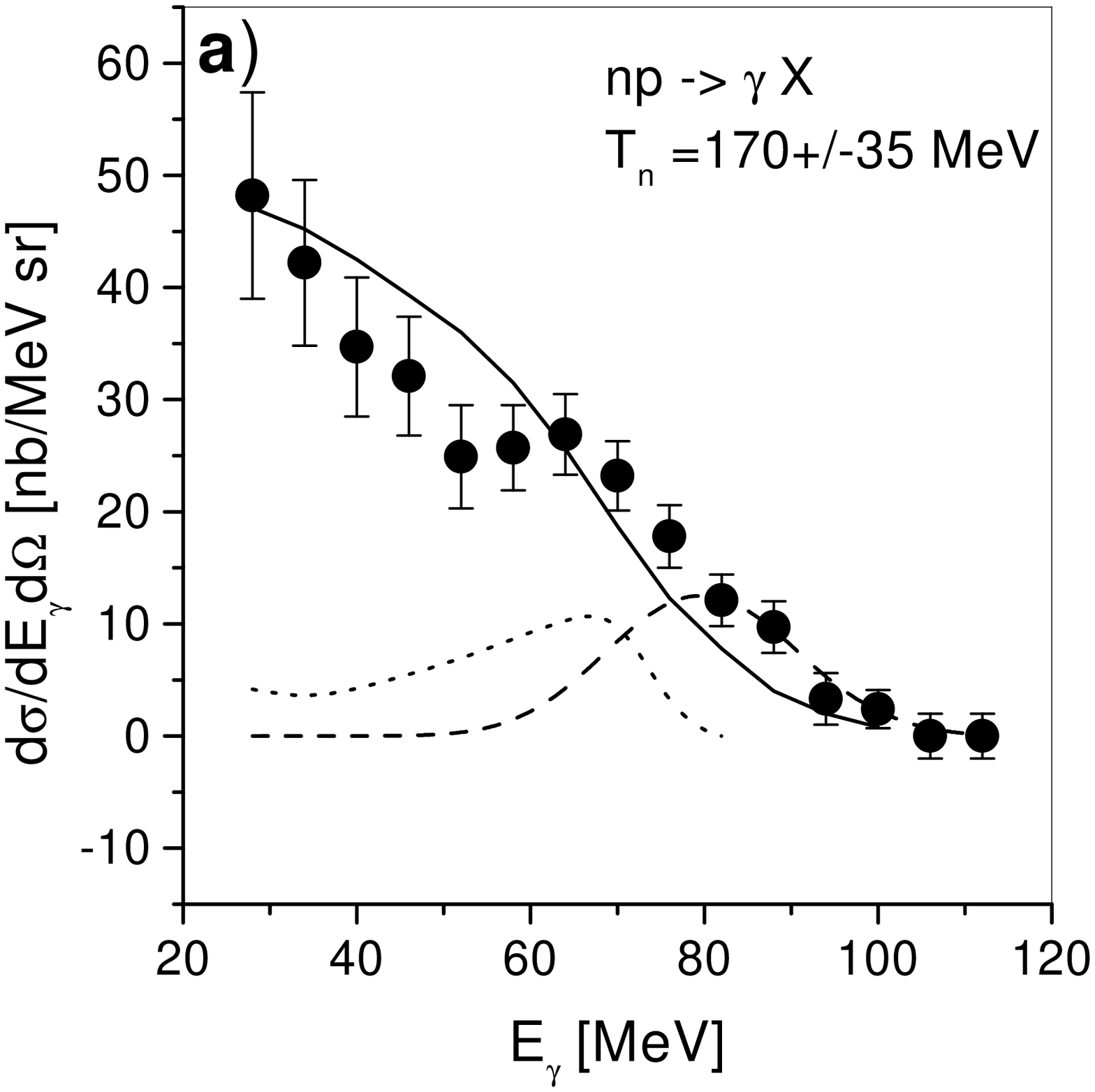}
\end{minipage}
\hspace{\fill}
%
\begin{minipage}[c]{75mm}
\includegraphics[width=70 mm]{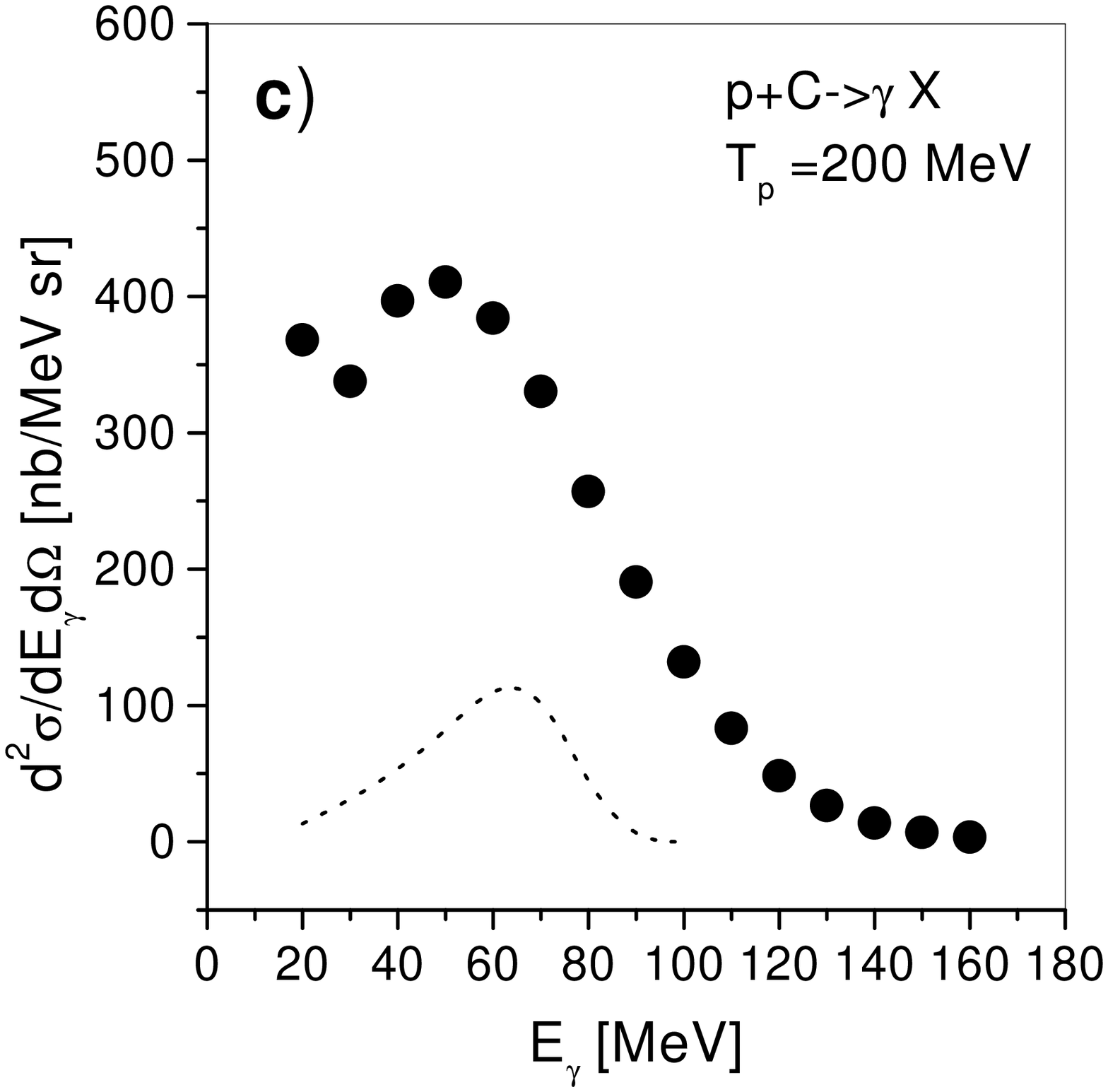}
\end{minipage}
\hspace{\fill}
\begin{minipage}[c]{75mm}
\includegraphics[width=70 mm]{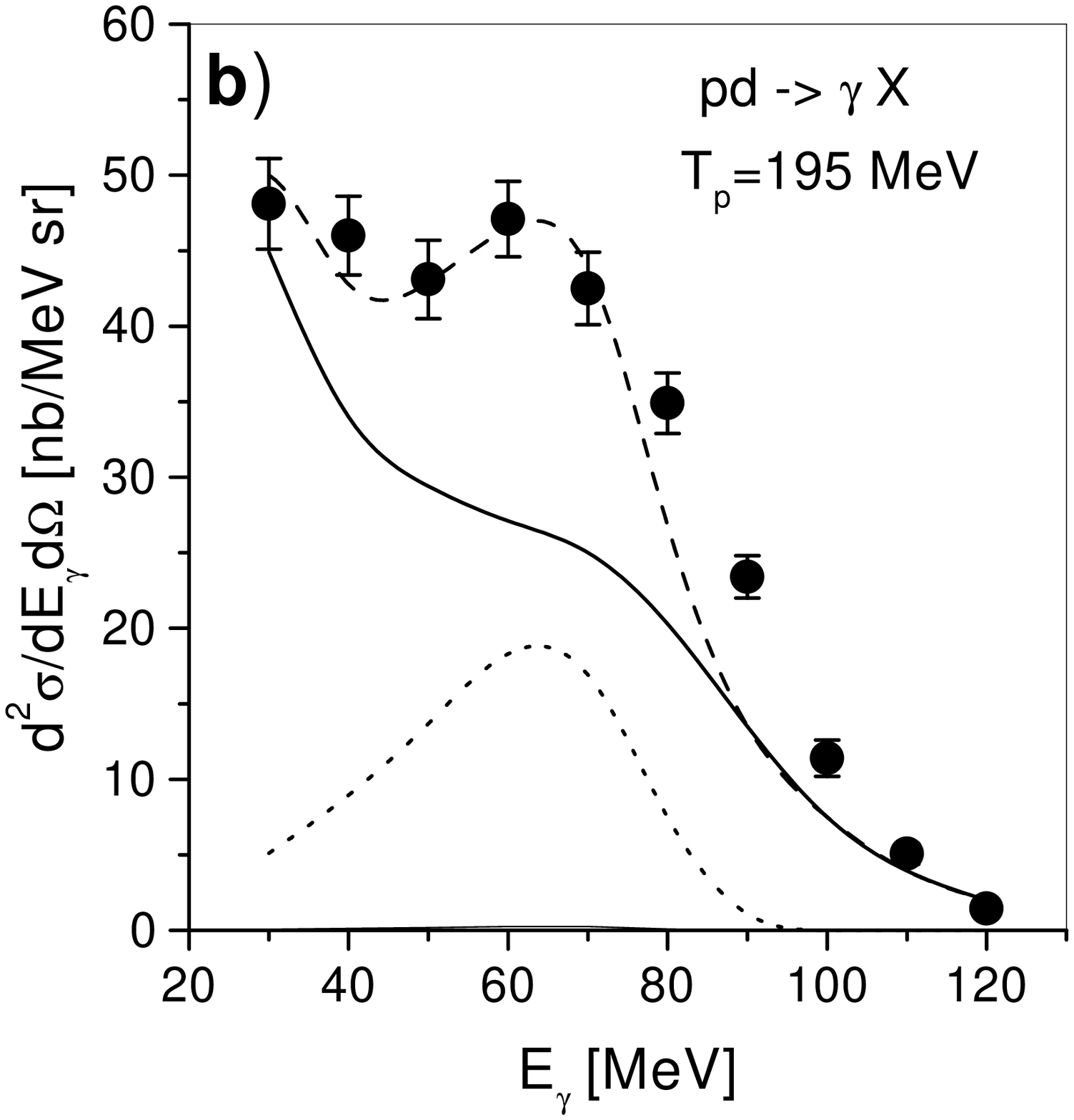}
\end{minipage}
\hspace{\fill}
\begin{minipage}[c]{80mm}
\vspace{-1.0 cm}
\caption{\small{Energy
spectra of photons emitted at $90^0_{lab}$ (solid circles).
a)The $np\gamma$ reaction at 170$\pm$35 MeV \cite{Malek}. The solid
line is the OBE model prediction, the dotted line shows the contribution
of the $d^\star_1$(1956) with the total production cross section
$\sigma_{tot}=4.3 \mu$b and the dashed line shows the contribution
of the $np\to d\gamma$ process with the same cross section.
b)The $pd\gamma$ reaction at 195 MeV \cite{Clayt}. The solid line is the
prediction \cite{Nak92}, the dotted line shows the $d^\star_1$(1956) contribution
with $\sigma_{tot}=9.4 \mu$b, and
dashed line shows the sum of the calculation \cite{Nak92} and $d^\star_1$(1956)
contribution. c) The $p+C \to \gamma X$ reaction \cite{Pinst}.
The dotted line shows the $d^\star_1$(1956) contribution
with $\sigma_{tot}=56.5 \mu$b}}
\end{minipage}
\end{figure}

\end{document}